\documentclass[preprint,12pt]{emulateapj}

\usepackage{apjfonts}
\usepackage{color}
\usepackage{natbib}
\usepackage{amssymb,amsmath}
\usepackage{graphicx}
\usepackage[colorlinks,urlcolor=red,citecolor=blue,linkcolor=cyan]{hyperref} 

\DeclareRobustCommand{\ion}[2]{%
\relax\ifmmode
\ifx\testbx\f@series
{\mathbf{#1\,\mathsc{#2}}}\else
{\mathrm{#1\,\mathsc{#2}}}\fi
\else\textup{#1\,{\mdseries\textsc{#2}}}%
\fi}

\shortauthors{Albrecht et al.} \shorttitle{EP\,Crucis}
\begin{document}

\title{The Banana project. IV.~Two aligned stellar rotation axes in the young
  eccentric binary system EP\,Crucis: primordial orientation and tidal alignment$^{\star}$}

\author{
 Simon Albrecht\altaffilmark{1}, 
Johny Setiawan\altaffilmark{2,3},
Guillermo Torres\altaffilmark{4},
Daniel C.\ Fabrycky\altaffilmark{5},
Joshua N.\ Winn\altaffilmark{1},
}

\altaffiltext{1}{Department of Physics, and Kavli Institute for
  Astrophysics and Space Research, Massachusetts Institute of
  Technology, Cambridge, MA 02139, USA}

\altaffiltext{2}{Max-Planck-Institut f\"{u}r Astronomie,
  K\"{o}nigstuhl 17, 69117 Heidelberg, Germany}

\altaffiltext{3}{Embassy of the Republic of Indonesia, Lehrter Str. 16-17, 10557 Berlin, Germany}

\altaffiltext{4}{Harvard-Smithsonian Center for Astrophysics,
  Cambridge, MA 02138, USA}

\altaffiltext{5}{Department of Astronomy and Astrophysics, University
  of California, Santa Cruz, Santa Cruz, CA 95064, USA}

\altaffiltext{$\star$}{Based on observations made with ESOs $2.2$\,m
  Telescopes at the La~Silla~Paranal~Observatory under program ID
  084.C-1008 ($12.5$\%) and under MPIA guaranteed time ($87.5$\%). }

\begin{abstract}

  With observations of the EP\,Cru system, we continue our series
  of measurements of spin-orbit angles in eclipsing binary
  star systems, the BANANA project (Binaries Are Not Always Neatly
  Aligned). We find a close alignment between the sky projections of
  the rotational and orbital  angular momentum vectors for both stars
  ($\beta_{\rm p}=  -1.8\pm1.6^{\circ}$ and $|\beta_{\rm s}|<17^{\circ}$). We also
  derive precise absolute dimensions and stellar ages for this system.
  The EP\,Cru and DI\,Her systems provide an interesting comparison: they
  have similar stellar types and orbital properties, but DI\,Her is
  younger and has major spin-orbit misalignments, raising the question
  of whether EP\,Cru also had a large misalignment at an earlier phase of
  evolution. We show that tidal dissipation is an unlikely
  explanation for the good alignment observed today, because
  realignment happens on the same timescale as spin-orbit
  synchronization, and the stars in EP\,Cru are far from
  synchronization (they are spinning 9 times too quickly). Therefore
  it seems that some binaries form with aligned axes, while other
  superficially similar binaries are formed with misaligned axes.

\end{abstract}


\section{Introduction}
\label{sec:intro}

One might expect star-planet and close binary star systems to have
well-aligned orbital and rotational angular momenta, since they
originate from the same portion of a molecular cloud. However, there
are also reasons to expect misaligned systems. Star formation is a
chaotic process, with accretion from different directions at different
times possibly leading to misalignment between the stellar and orbit
rotation axes \citep[e.g.,][]{bate2010,thies2011}. There are also
processes that could alter the stellar and orbital spin directions
after their formation. For example a third body orbiting a close pair
on a highly inclined orbit can introduce large oscillations in the
orbital inclination and eccentricity of the close pair
\citep{kozai1962}, thereby introducing large angles between the
stellar spins and orbital angular momentum of the close pair. Close
encounters and possible exchange of members in binary systems
\citep[e.g.,][]{gualandris2004} would leave, among other clues, a
fingerprint in the form of misalignment between the components. Tidal
forces will over time erase these clues, because dissipation will tend
to bring the axes into alignment while also synchronizing the
rotational and orbital periods
\citep[e.g.,][]{zahn1977,hut1981,eggleton2001}. Thus the degree of
alignment between the stellar rotation axes and the orbital axis
depends on its particular history of formation and evolution.
Therefore measurements of stellar obliquities -- the angle between
stellar equator and orbital plane -- allow us to test theories of
formation and evolution in close star-planet and star-star systems.

For example the formation of star-star systems with orbital distances
of only a few stellar radii is not completely understood. It seems
unlikely that the stars formed at these orbital distances because they
would have overlapped during their pre-main sequence phase, when they
had larger sizes. Therefore the orbital distance likely decreased
after formation. A possible mechanism is KCTF -- Kozai Cycles with
Tidal Friction \citep{eggleton2001,fabrycky2007}, which requires a
third body on a wide orbit around the close pair. \cite{tokovinin2006}
found that $96\%$ of binary stars with orbital periods less than
$3$~days have a third companion on a wide orbit while only $34\%$ of
binaries with orbital periods larger than $12$~days have a third
companion.  Additional evidence for KCTF would be a misalignment
between the stellar spin axes and the orbital spin, assuming that
close binaries have aligned axes at birth, and that tides have not had
enough time to align the spin axes.  Thus measurements of stellar
obliquities in close binary systems together with a good understanding
of tidal dissipation in these systems might lead to a better
understanding of binary formation.

For the case of close star-planet (hot-Jupiter) systems such an
approach has been fruitful. Hot-Jupiters are thought to have formed
much further from the star than their current orbital distances,
mainly because not enough material would have been available so close
to the star. Different processes which could have transported the
planet inward would lead to different spin-orbit angles, and indeed
systems with both small and large spin-orbit angles have been found
\citep[see, e.g.,][]{winn2005,hebrard2008, johnson2009,
  albrecht2012,brown2012}. In addition \cite{winn2010} and
\cite{albrecht2012b} presented evidence that all hot-Jupiter systems
once had high obliquities, and that tides are responsible for the
frequently observed low obliquities. This suggests that the inward
migration of hot Jupiters involves changes of the orbital planes of the
planets.

With the BANANA project (Binaries Are Not Always Neatly Aligned) we
aim to get a better understanding of the formation of close binaries
as well as their tidal spin evolution. Here we study the EP\,Cru
binary system. This is the fourth system which we study as part of the
BANANA project \citep[][Papers
I--III]{albrecht2007,albrecht2009,albrecht2011}. We also refer the
reader to \cite{triaud2012}, for a description of a similar project by
other investigators. While most of the stars in our sample are of
early spectral types, their EBLM project focuses on eclipsing systems
harboring low-mass stars.

EP\,Cru was only recently characterized by \cite{clausen2007}.
Table~\ref{tab:nsv5783_overview} gives some system parameters. We
selected this system because \cite{clausen2007} found it to be similar
to DI\,Her, for which we already found the spin-orbit angles to be
very large \citep{albrecht2009}. In particular the orbital parameters,
the stellar masses, and the projected stellar rotation speeds ($v\sin
i_{\star}$), are similar in these two systems. Here $v$ indicates the
equatorial rotation speed and $i_{\star}$ the inclination of the
stellar rotation axis along the line of sight (LOS). There is one
important difference between the two systems: the age of the stars.
\cite{clausen2007} estimated an age of $\approx50$~Myr for the two
stars in the EP\,Cru system while DI\,Her is essentially a Zero Age
Main Sequence (ZAMS) system with an estimated age of $4.5\pm2.5$~Myr
\citep{claret2010}. Therefore by studying EP\,Cru we have the
opportunity to learn if according to our current understanding of
binary evolution one system is simply an older version of the other,
or if they had different childhoods altogether.

\begin{table}[t]
 \caption{General data on EP\,Cru}
 \label{tab:nsv5783_overview}
 \begin{center}
 \smallskip
     \begin{tabular}{l l l}
	\hline
	\hline
	\noalign{\smallskip}
        HD & 109724  &  \\ 
        NSV &  5783    &  \\ 
	R.A.$_{\rm J2000}$ & $12^{\rm h}37^{\rm m}17^{\rm s}$&\tablenotemark{$\dagger$} \\
	Dec.$_{\rm J2000}$ & $-56^{\circ}47^{\prime}17^{\prime\prime}$&\tablenotemark{$\dagger$}  \\
        Distance  & $1.0 (1)$ kpc &\tablenotemark{$\star$}\\
	V$_{\mbox{max}}$  & $8.7$\,mag& \tablenotemark{$\star$} \\
	Sp.\ Type  & B5V+ B5V &\tablenotemark{$\star$} \\
	Orbital period  & $11\fd08$&\tablenotemark{$\star$}\\
        Eccentricity  & $0.19$&\tablenotemark{$\star$} \\
        $R_{\rm p}$ & $3.6(3) R_{\odot}$&\tablenotemark{$\star$} \\
        $R_{\rm s}$ & $3.5(2) R_{\odot}$&\tablenotemark{$\star$} \\
        $M_{\rm p}$ & $5(1) M_{\odot}$&\tablenotemark{$\star$} \\
        $M_{\rm s}$  & $4.7(1.1) M_{\odot}$&\tablenotemark{$\star$} \\
        $T_{\rm eff\,p}$ & $15\,700(500)$K&\tablenotemark{$\star$} \\
        $T_{\rm eff\,s}$   & $15\,400(500)$K&\tablenotemark{$\star$} \\
	\noalign{\smallskip}
        \noalign{\smallskip}
	\hline
	\noalign{\smallskip}
        \noalign{\smallskip}
        $\dagger$Data from \cite{esa1997}\\
        $\star$Data from \cite{clausen2007}\\
     \end{tabular}
     \tablecomments{$R_{\rm p}$ denotes the radius of the primary
       component and $R_{\rm s}$ the radius of the secondary
       component. $M_{\rm p}$ and $M_{\rm p}$ denote the masses
       $T_{\rm eff\,s}$ and $T_{\rm eff\,s}$ denote the effective
       temperatures.} 
\end{center} 
\end{table}

The plan of this paper is as follows. In the following section we
describe our observations. The analysis of the spectroscopic
observations during eclipses and out of eclipses is presented in
Section~\ref{sec:analysis}. We present the results on the absolute
dimensions, the orientations of the stellar rotation axes, and the
derived age in Section~\ref{sec:results}. For the remainder of the
paper we focus on the interpretation of the obliquity measurements
before we summarize our findings in the conclusions.

\section{Spectroscopic Observations}
\label{sec:obs}

\begin{table}
  \begin{center}
    \caption{Observation Log for EP\,Cru\label{tab:log_nsv5783}}
    \smallskip 
    {\footnotesize
    \begin{tabular}{c c c c }
      \tableline\tableline
      \noalign{\smallskip}
        Obs. Mid Time  & Phase &   Eclipse & S/N \\
        (BJD$_{\rm TDB}$) &         &                &       \\
      \noalign{\smallskip}
      \hline
      \noalign{\smallskip}
  $2455192.85669$  & $0.70$ &  --  &  $ 142$  \\
  $2455193.83164$  & $0.78$ &  --  &  $  97$  \\
  $2455194.69509$  & $0.86$ & sec  &  $  77$  \\
  $2455194.83360$  & $0.87$ &  --  &  $ 107$  \\
  $2455196.85095$  & $0.06$ &  --  &  $ 107$  \\
  $2455197.86201$  & $0.15$ &  --  &  $ 115$  \\
  $2455309.59755$  & $0.24$ & pri  &  $  84$  \\
  $2455309.63874$  & $0.24$ & pri  &  $  86$  \\
  $2455309.68141$  & $0.24$ & pri  &  $  91$  \\
  $2455309.71894$  & $0.25$ & pri  &  $  85$  \\
  $2455309.75526$  & $0.25$ & pri  &  $  89$  \\
  $2455309.80483$  & $0.25$ & pri  &  $  81$  \\
  $2455309.84282$  & $0.26$ &  --  &  $  71$  \\
  $2455309.86946$  & $0.26$ &  --  &  $  83$  \\
  $2455310.46906$  & $0.31$ &  --  &  $  73$  \\
  $2455311.46870$  & $0.40$ &  --  &  $  96$  \\
  $2455313.48255$  & $0.59$ &  --  &  $  94$  \\
  $2455314.57372$  & $0.68$ &  --  &  $  67$  \\
  $2455338.45972$  & $0.84$ & sec  &  $  60$  \\
  $2455338.48823$  & $0.84$ & sec  &  $  82$  \\
  $2455338.52361$  & $0.85$ & sec  &  $  94$  \\
  $2455338.57536$  & $0.85$ & sec  &  $  97$  \\
  $2455338.61565$  & $0.85$ & sec  &  $  62$  \\
  $2455338.65529$  & $0.86$ & sec  &  $  64$  \\
  $2455338.69515$  & $0.86$ & sec  &  $ 103$  \\
  $2455338.72942$  & $0.86$ & sec  &  $  54$  \\
  $2455340.44257$  & $0.02$ &  --  &  $  78$  \\
  $2455341.55552$  & $0.12$ &  --  &  $ 140$  \\
  $2455342.55950$  & $0.21$ & pri  &  $ 120$  \\
  $2455353.46096$  & $0.19$ &  --  &  $ 130$  \\
  $2455353.50079$  & $0.20$ &  --  &  $ 114$  \\
  $2455353.52531$  & $0.20$ &  --  &  $ 144$  \\
  $2455353.57109$  & $0.20$ & pri  &  $ 123$  \\
  $2455353.61561$  & $0.21$ & pri  &  $ 137$  \\
  $2455353.65050$  & $0.21$ & pri  &  $ 128$  \\
  $2455360.46345$  & $0.83$ & sec  &  $  83$  \\
  $2455360.53196$  & $0.83$ & sec  &  $  89$  \\
  $2455360.60002$  & $0.84$ & sec  &  $  53$  \\
  $2455360.64101$  & $0.84$ & sec  &  $  44$  \\
  $2455361.49041$  & $0.92$ &  --  &  $ 103$  \\
  $2455362.49566$  & $0.01$ &  --  &  $  88$  \\
  $2455628.85227$  & $0.06$ &  --  &  $  85$  \\
  $2455630.66346$  & $0.22$ & pri  &  $  82$  \\
  $2455630.72346$  & $0.22$ & pri  &  $  92$  \\
  $2455630.78472$  & $0.23$ & pri  &  $  80$  \\
  $2455630.85156$  & $0.24$ & pri  &  $  91$  \\
  $2455632.71659$  & $0.40$ &  --  &  $  83$  \\
  $2455635.68609$  & $0.67$ &  --  &  $  81$  \\
      \noalign{\smallskip}
      \tableline
      \noalign{\smallskip}
      \noalign{\smallskip}
    \end{tabular}
    \tablecomments{The phase is defined such that phase 0 corresponds to
      periastron. In column 3 'pri' indicates that the observation
      was taken during a primary eclipse and 'sec' indicates that the
      observation was obtained during an ongoing secondary eclipse.} 
  }
  \end{center}
\end{table}

\begin{figure*}
  \begin{center}
\includegraphics[width=16cm]{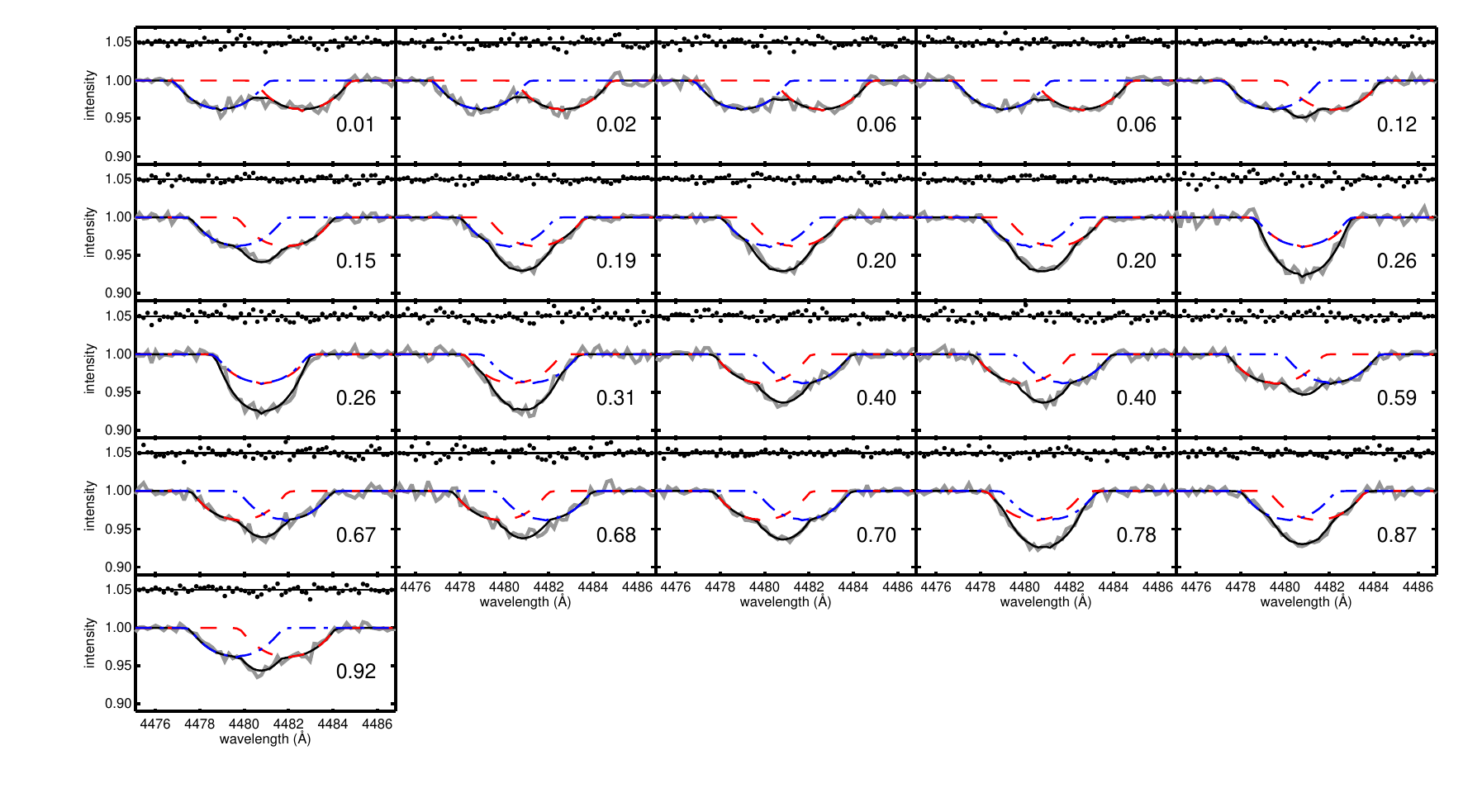}
\caption {\label{fig:nsv5783_orbit} {\bf Spectra of EP\,Cru obtained
    at  different orbital phases.}  Each panel shows spectra of both
  stars in the spectral region around the \ion{Mg}{ii} line. The
  orbital phase of the observation is indicated by the number in
  each panel, with phase 0 occurring at periastron passage. The gray
  solid lines represent the data and the (red) dashed and (blue)
  dash-dotted lines are the simulated absorption lines of the primary
  and secondary, respectively. The black line is the best fitting
  model. The dots around the line at a flux level of 1.05 represent
  the differences between the data and the model.}
  \end{center}
\end{figure*}

We observed the EP\,Cru system with the FEROS spectrograph
\citep{kaufer1999} on the $2.2$~m telescope at ESO's La\,Silla
observatory. We obtained 48 observations on multiple nights between
2009 December and 2011 March. Table~\ref{tab:log_nsv5783} gives an
observation log. The observations had a typical integration time of
10\,min. On each night, ThAr exposures were taken to calculate a
wavelength solution and monitor any changes in the spectrograph. For
all observations we used the MIDAS FEROS package installed on the
observatory computers to reduce the raw 2D CCD images and to obtain
stellar flux as a function of wavelength. The uncertainty in the
wavelength solution, expressed in velocity, is few m\,s$^{-1}$, and is
negligible for our purposes. The resulting spectra have a resolution
of $\approx50\,000$ around $4481$~\AA{} (the wavelength area of the
spectra we analyze). We corrected for the radial-velocity (RV) of the
observatory, performed initial flat fielding with the nightly blaze
function, and flagged and omitted bad pixels.

\section{Analysis}
\label{sec:analysis}

In this section we outline the analysis of the spectra with the aim of
deriving absolute dimensions of the system and learning about the
projected obliquities of both stars via measurements of the
Rossiter-McLaughlin (RM) effect, which occurs during eclipses. We
describe which part of spectrum we analyze and briefly introduce the
model to which we compare the data and the algorithm used to extract
system parameters. Our approach for EP\,Cru is similar to the approach
employed in Papers~I--III.

\paragraph{Spectral region}

We focus on the \ion{Mg}{ii} line at $4481$~\AA{}, as this line is
relatively deep and chiefly broadened by stellar rotation. It is
located in the red wing of the pressure-broadened \ion{He}{i} line at
$4471$~\AA{}. While this line might also be included in the analysis
\citep{albrecht2011}, we decided here to exclude it as there is enough
signal in the \ion{Mg}{ii} line and modeling the pressure broadening
in the \ion{He}{i} line represents an additional complication. Thus we
fitted a Lorentzian model to the encroaching wing of the \ion{He}{i} line and
subtracted it before modeling the \ion{Mg}{ii} line. For this fit we
used the spectral regions $4472$ -- $4476$~\AA{} and
$4486$ -- $4498$~\AA{}, thereby avoiding the influence of the
\ion{Mg}{ii} line. Each spectrum was binned to a resolution of about
12~km\,s$^{-1}$, to speed up subsequent computations. Because the
stellar rotation speeds are an order of magnitude larger, there is no
significant loss of information due to the binning; we verified this
by experimenting with higher resolutions.

\begin{figure*}
  \begin{center}
    \includegraphics[width=16.cm]{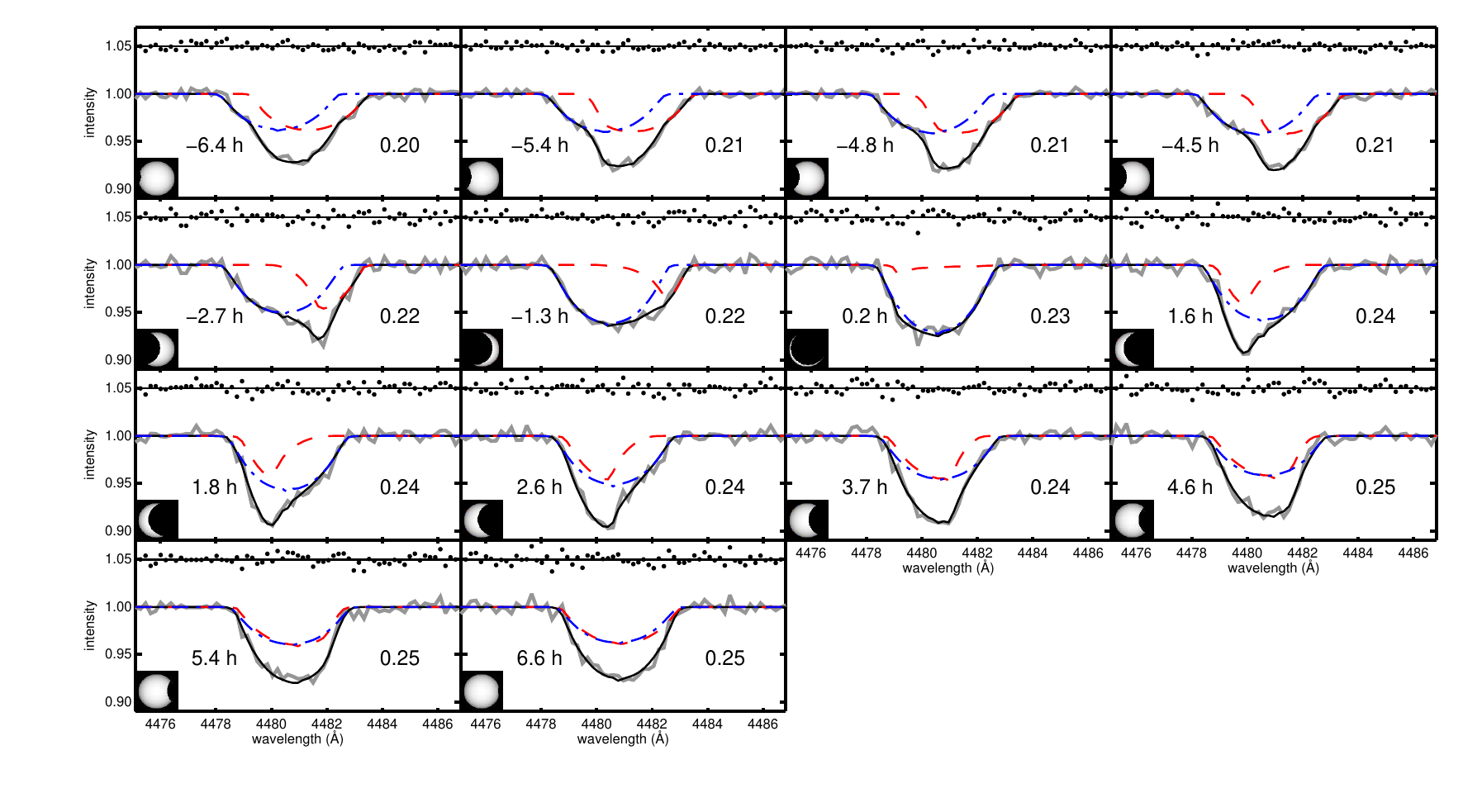}
    \caption {\label{fig:nsv5783_primary} {\bf Spectra of EP\,Cru
        obtained during primary eclipse.} Similar to
      Figure~\ref{fig:nsv5783_orbit}, but this time for observations
      obtained during primary eclipses. The numbers on the left side
      of each panel indicate the  observation mid-exposure times
      relative to the time of minimum  light, in hours. Each  inset
      shows an illustration of the eclipse phase of the
      background star. }
  \end{center}
\end{figure*}

\begin{figure*}
  \begin{center}
    \includegraphics[width=16.cm]{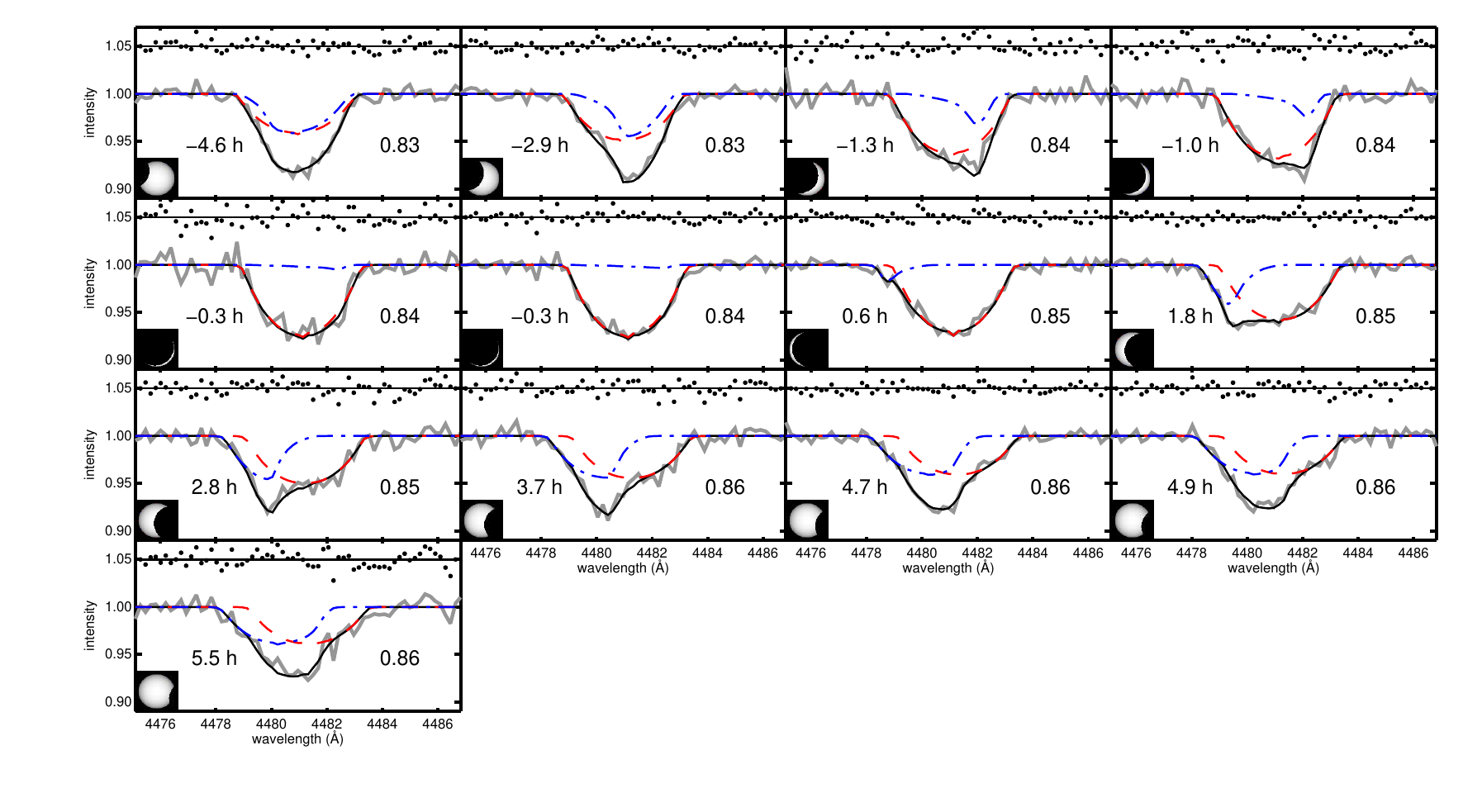}
    \caption {\label{fig:nsv5783_secondary} {\bf Spectra of EP\,Cru
        obtained during secondary eclipse.} Similar to
      Figure~\ref{fig:nsv5783_primary}, but for spectra obtained
      during secondary eclipses. }
  \end{center}
\end{figure*}

\paragraph{Model}
 
The measured spectra show absorption lines of both stars in the
system. Before, during, and after eclipses the RVs of both stars are
similar, leading to a substantial overlap of the two absorption lines.
Hence, light emitted from both stars has to be accounted for when
analyzing the RM effect. We used the numerical code from
\cite{albrecht2007} which simulates the spectra of both stars in a
system.

\begin{table*}
 \begin{center}
  \caption{Results for the EP\,Cru System.\label{tab:results}}
    \smallskip 
       \begin{tabular}{l  r@{$\pm$}l r@{$\pm$}l   }
          \tableline\tableline
          \noalign{\smallskip}
          Parameter &  \multicolumn{2}{c}{ This Work}  &    \multicolumn{2}{c}{Literature Values} \\
          \noalign{\smallskip}   
          \hline
          \noalign{\smallskip}
          \multicolumn{5}{c}{Orbital parameters} \\
          \noalign{\smallskip}
          \hline
          \noalign{\smallskip}
          Time of primary minimum, T$_{\rm min, I}$ (BJD$-$2\,400\,000)  &  46181.6928&0.0038  & 46181.7068&0.0003$^1$   \\
          Period, $P$ (days)                                &   11.0774707&0.0000043   & 11.0774701&0.0000042$^1$$^{\rm c}$  \\
         $\sqrt{e} \cos \omega$                         &   -0.104&0.006  &   \multicolumn{2}{c}{}          \\
          $\sqrt{e} \sin \omega$                         &     0.4202&0.0011   &  \multicolumn{2}{c}{}         \\   
          $ e \cos \omega$                                  &    0.18187&0.00034   & 0.18162&0.00007$^1$       \\
          $ e \sin \omega$                                  &   -0.0450&0.0026   &   -0.0475& 0.0020$^1$      \\   
          Eccentricity, $e$                                  &   0.1874&0.0005  &   0.1877&0.0005$^1$$^{\rm c}$       \\
          Argument of periastron, $\omega$ (deg)   & 346.1&0.7  &  345.4&0.6$^1$        \\   
          Cosine orbital inclination, $\cos i_{o}$     &  0.0052&0.0014      &   \multicolumn{2}{c}{}      \\   
          Orbital inclination, $i_{o}$ (deg)                  & 89.70&0.08     &     86.97&0.09$^1$$^{\rm c}$\\%\footnote{They have a typo in thier Table 8.}      \\   
         Velocity semi-amplitude primary, $K_{\rm p}$ (km\,s$^{-1}$)     &     102.2 &1.5    & 100.9&1.3$\pm10$$^1$          \\
          Velocity semi-amplitude secondary, $K_{\rm s}$ (km\,s$^{-1}$)    &    106.2&1.4  &      105.9&3.5$\pm10$$^1$     \\
          Velocity offset, $\gamma_{p}$ (km\,s$^{-1}$)   &     $-26.3$&0.6   &     \multicolumn{2}{c}{$-33$$^1$}         \\
          Velocity offset, $\gamma_{p}$ (km\,s$^{-1}$)   &     $-27.2$&0.5   &     \multicolumn{2}{c}{$-33$$^1$}         \\
          Orbital semi-major axis,    $a$ ($R_{\odot}$)      &    44.83&0.37     &   44.6&4.2$^1$     \\
          \noalign{\smallskip}
          \hline
          \noalign{\smallskip}
          \multicolumn{5}{c}{Stellar  parameters} \\
          \noalign{\smallskip}
          \hline 
          \noalign{\smallskip}
          Light ratio, $L_{\rm s}/L_{\rm p}$  @  4480\,\AA{}     &   0.8972 & 0.0020   &   0.8972&0.0020$^1$$^{\rm c}$\\%\footnote{Their value in the $b$ band}     \\
          Fractional radius primary, $r_{\rm p}$         &    0.0801&0.0005    &       0.0810&0.0006$^1$$^{\rm c}$   \\
          Fractional radius secondary, $r_{\rm s}$    &    0.0779&0.0004   &      0.0782&0.0006$^1$$^{\rm c}$      \\
          $u1_{\rm i} $+$u2_{\rm i} $                               &    0.50&0.05         &      \multicolumn{2}{c}{(0.4+0.1)$\pm$0.1$^2$$^{\rm c}$}              \\  
          Macro-turbulence parameter,  $\zeta_{\rm p}$ (km\,s$^{-1}$)    &    22.3&1.7                              & \multicolumn{2}{c}{}          \\
          Projected rotation speed primary,     $v \sin i_{\rm p}$ (km\,s$^{-1}$)     &    141.4&1.2$\pm$5   & \multicolumn{2}{c}{150$^1$}   \\
          Projected rotation speed secondary,   $v \sin i_{\rm s}$ (km\,s$^{-1}$)   &    137.8&1.1$\pm$5   & \multicolumn{2}{c}{150$^1$}      \\
          Projected spin-orbit angle primary,   $\beta_{\rm p}$ ($^{\circ}$)     &    $-1.8$&$1.6$   & \multicolumn{2}{c}{}       \\ 
          Projected spin-orbit angle secondary,   $\beta_{\rm s}$  ($^{\circ}$)   &    \multicolumn{2}{c}{$<17$$^{\rm a}$}     & \multicolumn{2}{c}{}       \\ 
          Primary mass,     $M_{\rm p} $ ($M_{\odot}$)   &  5.02&0.13$^{\rm b}$  & 4.95&1.06$^1$  \\
          Secondary mass, $M_{\rm s} $ ($M_{\odot}$)   &  4.83&0.13$^{\rm b}$ &  4.72&1.03$^1$      \\
          Primary radius,     $R_{\rm p} $ ($R_{\odot}$)   &  3.590&0.035$^{\rm b}$ & 3.61&0.25$^1$ \\
          Secondary radius, $R_{\rm s} $ ($R_{\odot}$)   &  3.495&0.034$^{\rm b}$ & 3.48&0.24$^1$     \\     
          Primary $\log g_{\rm p}$ (cgs)                      &  4.028&0.008 & 4.02&0.11$^1$      \\
          Secondary $\log g_{\rm s}$ (cgs)                  &  4.035&0.008 & 4.03&0.11$^1$   \\
          \noalign{\smallskip}
          \tableline
           \noalign{\smallskip}
           \noalign{\smallskip}
            \multicolumn{5}{l}{{\sc Notes} ---}\\
           \multicolumn{5}{l}{$^{\rm a}$ We prefer this larger
             uncertainty interval over the statistical uncertainty of $-13\pm4^{\circ}$ (Section~\ref{sec:rot}).}\\
            \multicolumn{5}{l}{$^{\rm b}$ A solar radius of $ 6.9566\cdot10^{8}$~m was used.}\\
        \multicolumn{5}{l}{$^{\rm c}$ Value was used as prior.}\\
        \noalign{\smallskip}
        \multicolumn{5}{l}{{\sc References} ---}\\
        \multicolumn{5}{l}{(1) \cite{clausen2007} (2) \cite{claret2000}}
        \end{tabular}
     \end{center}
  \end{table*}

\begin{figure*}
  \begin{center}
    \includegraphics[width=18cm]{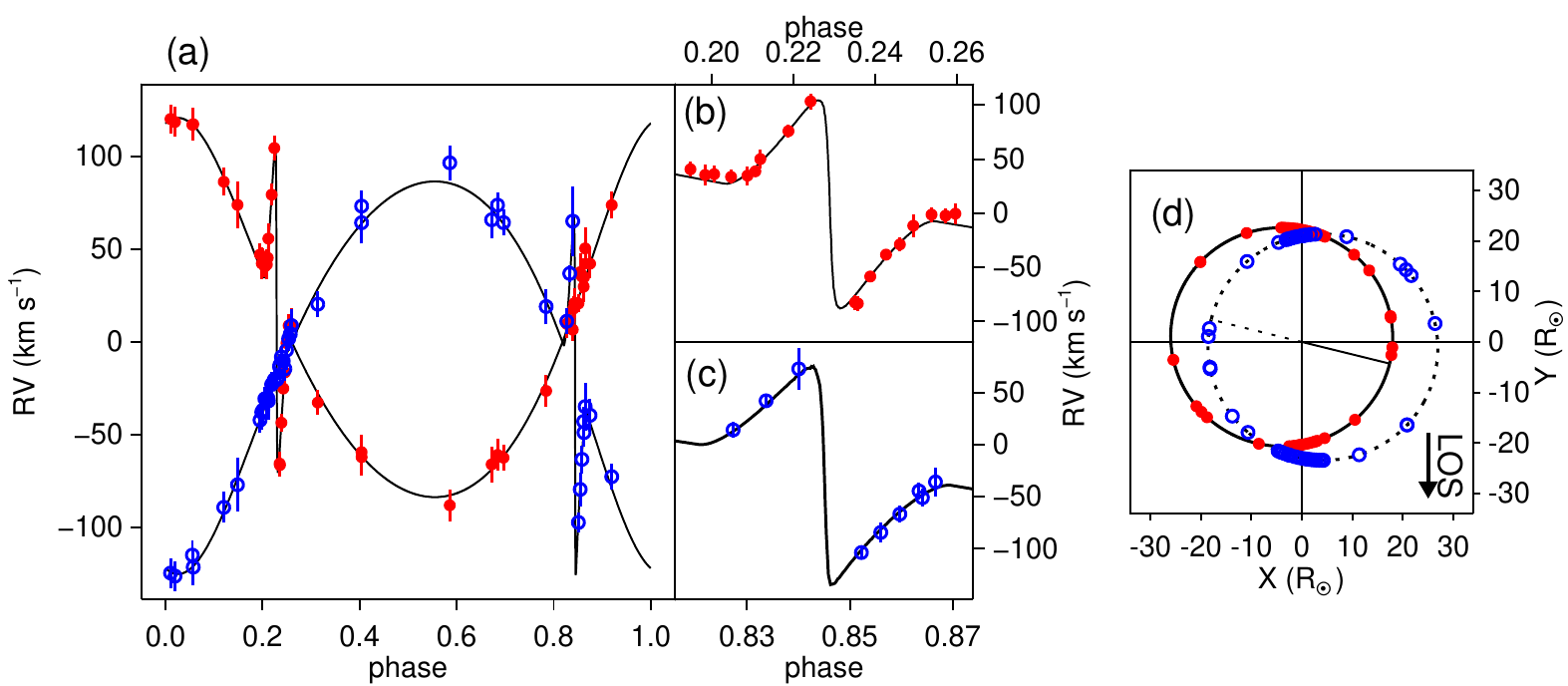}
    \caption {\label{fig:nsv5783_rvs} {\bf Orbit of EP\,Cru.} {\bf
        (a)} The apparent radial velocities (RV) in the EP\,Cru system
      for both stars as a function of orbital phase, defined such that
      phase zero occurs at periastron passage. The (red) filled circles indicate
      measured RVs of the primary and the open (blue) circles RVs of
      the secondary. Due to the small amount of light received from
      the eclipsed star during mid-eclipses it was not possible to
      assign RVs to the background stars for observations obtained
      within $1$~hr of mid-eclipse. The RVs shown here were not
      used in the analysis; they are displayed only for illustrative
      purposes. {\bf (b)} Zoom-in on orbital phases close to primary
      eclipse. Only RVs of the primary are shown. {\bf (c)} The same
      as {\bf b}, but this time for the secondary eclipse. {\bf (d)} A
      pole on view of the orbit of both stars. The line of sight (LOS)
      towards earth is indicated.}
  \end{center}
\end{figure*}

The stellar disks are discretized with $\sim30,000$ pixels in a
Cartesian coordinate system. We assume the stars to be spherical
because they are well separated with rotation speeds much slower than
the breakup velocity; \cite{clausen2007} estimates an oblateness of
about $0.0008$. We further assume uniform rotation and quadratic
limb darkening. Stellar surface velocity fields are parameterized
adopting the macro-turbulence model by \cite{gray2005}.\footnote{Here
  we do not need to take the point spread function (PSF) of the
  spectrograph into account as our binning ($12$~km\,s$^{-1}$) is
  larger than the width of the PSF ($6$~km\,s$^{-1}$).} The
coordinates of both stars projected on the sky are calculated and
light from visible parts of the stellar hemispheres is integrated. The
resulting absorption line kernels are shifted in wavelength 
corresponding to the line they represent and weighted according to the
light contribution of the respective star.

\paragraph{Parameter choices}

Having the model in place we can now learn about the EP\,Cru system by
specifying a number of parameters. The Keplerian orbit of the two
stars can be described with the following $6$ parameters: The orbital
period ($P$), a specific time of minimum light during primary eclipse
($T_{\rm min, I}$), the orbital eccentricity ($e$), the argument of
periastron ($\omega$), and the velocity semi-amplitudes of the primary
and secondary stars ($K_{\rm i}$). Here the subscript 'i' stands for
either 'p' indicating the primary star or 's' indicating the secondary
(slightly less massive) star. In addition, velocity offsets
($\gamma_{\rm i}$) are needed. For $e$ and $\omega$ we use the
stepping parameters $\sqrt{e} \cos \omega$ and $\sqrt{e} \sin \omega$,
as they are less correlated than $e$ and $\omega$ themselves. The
results from the photometric study by \cite{clausen2007} can be used
to constrain some of these orbital parameters. However the photometry
used by \cite{clausen2007} was gathered about 20 years ago and the
system is expected to have an apsidal motion period of a few thousand
years. The change in $\omega$ could be of order $1^\circ$ over the
last two decades. In addition the apsidal motion is not measured yet
and we cannot calculate it from the known system parameters as it
depends on the true stellar obliquity and not only the sky projection
\citep{shakura1985,albrecht2009}. Thus we do not use the photometric
values on $T_{\rm min I,}$ and $\omega$ as priors and we only use the
measurements of $P$ and $e$ by \cite{clausen2007} as prior
constraints. We revisit this subject in Section~\ref{sec:apsidal}.

Additional parameters are needed to describe the projected equatorial
rotation speeds ($v \sin i_{\rm i}$), the Gaussian width of the
macro-turbulence ($\zeta_{\rm i}$), and the parameters of greatest
interest for this study, the sky-projected spin-orbit angles
($\beta_{\rm i}$). The angle is defined according to the convention of
\citet{hosokawa1953}.

The photometric character of the eclipses are specified by another set
of parameters: the light ratio between the two stars at the wavelength
of interest ($L_{\rm s}/L_{\rm p}$ at 4480\,\AA{}), the quadratic limb
darkening parameters ($u1_{\rm i}$ and $u2_{\rm i}$), the fractional
radii of the stars ($r_{\rm i}$), and the orbital inclination ($i_o$),
for which we step in $\cos i_o$. For the fractional radii and the
orbital inclination we use prior information from \cite{clausen2007}.
For $L_{\rm s}/L_{\rm p}$ we use their results in the {\it b} band. To
constrain the limb darkening parameters we used the
'jktld'\footnote{{\tt
    http://www.astro.keele.ac.uk/jkt/codes/jktld.html}} tool to query
the ATLAS atmospheres \citep{claret2000} and placed a Gaussian prior
on $u1_{\rm i}+u2_{\rm i}$ with a width of 0.1 and held the difference
$u1_{\rm i}-u2_{\rm i}$ fixed at the tabulated value.

An additional parameter is needed for each star to describe the
relative depth of the \ion{Mg}{ii} lines. The \ion{Mg}{ii} line
consists of a doublet, given the close spacing ($ 0.2$\,\AA{}) we
model it here as single line.

The two components in the EP\,Cru system are very similar to each
other (see Table~\ref{tab:nsv5783_overview}). We therefore decided to
use the same limb darkening parameters and macro-turbulence
velocities, for both stars, thereby reducing the number of free
parameters to 20. Of these, 7 are further constrained by Gaussian
priors as explained above. Table \ref{tab:results} summarizes all of
the prior constraints.

There is always a small residual uncertainty in the initial
normalization of the spectra. To propagate this into the uncertainty
intervals of the final parameters we added for each of the $48$
observation $3$ free parameters which describe a quadratic function
used to normalize the continuum level. The values of the normalization
parameters were optimized using a separate 3-parameter minimization
for each observation, each time a set of system parameters is
evaluated. This process is similar to the ``Hyperplane Least Squares''
method that was described and tested by \cite{bakos2010}.

\paragraph{Parameter estimation} A MCMC code was used to obtain
uncertainty intervals. The chains consisted of $0.5$ million
calculations of $\chi^2$. The results reported below are the median
values of the posterior distribution and the uncertainties intervals
are the values which exclude $15.85$~\% of the values at each side of
the posterior and encompassing $68.3$~\% of all values.

\newpage
\section{Results}
\label{sec:results}

The results for the model parameters are given in
Table~\ref{tab:results}. Figure~\ref{fig:nsv5783_orbit} shows the
spectra in the vicinity of the \ion{Mg}{ii} line and the corresponding
model for the out-of-eclipse observations.
Figures~\ref{fig:nsv5783_primary} and \ref{fig:nsv5783_secondary} show
the same for the spectra obtained during primary and secondary
eclipses. The apparent radial velocities in the EP\,Cru system are
shown in Figure~\ref{fig:nsv5783_rvs} as well as a pole-on view of the
orbit.

\begin{figure}
  \begin{center}
\includegraphics[width=8cm]{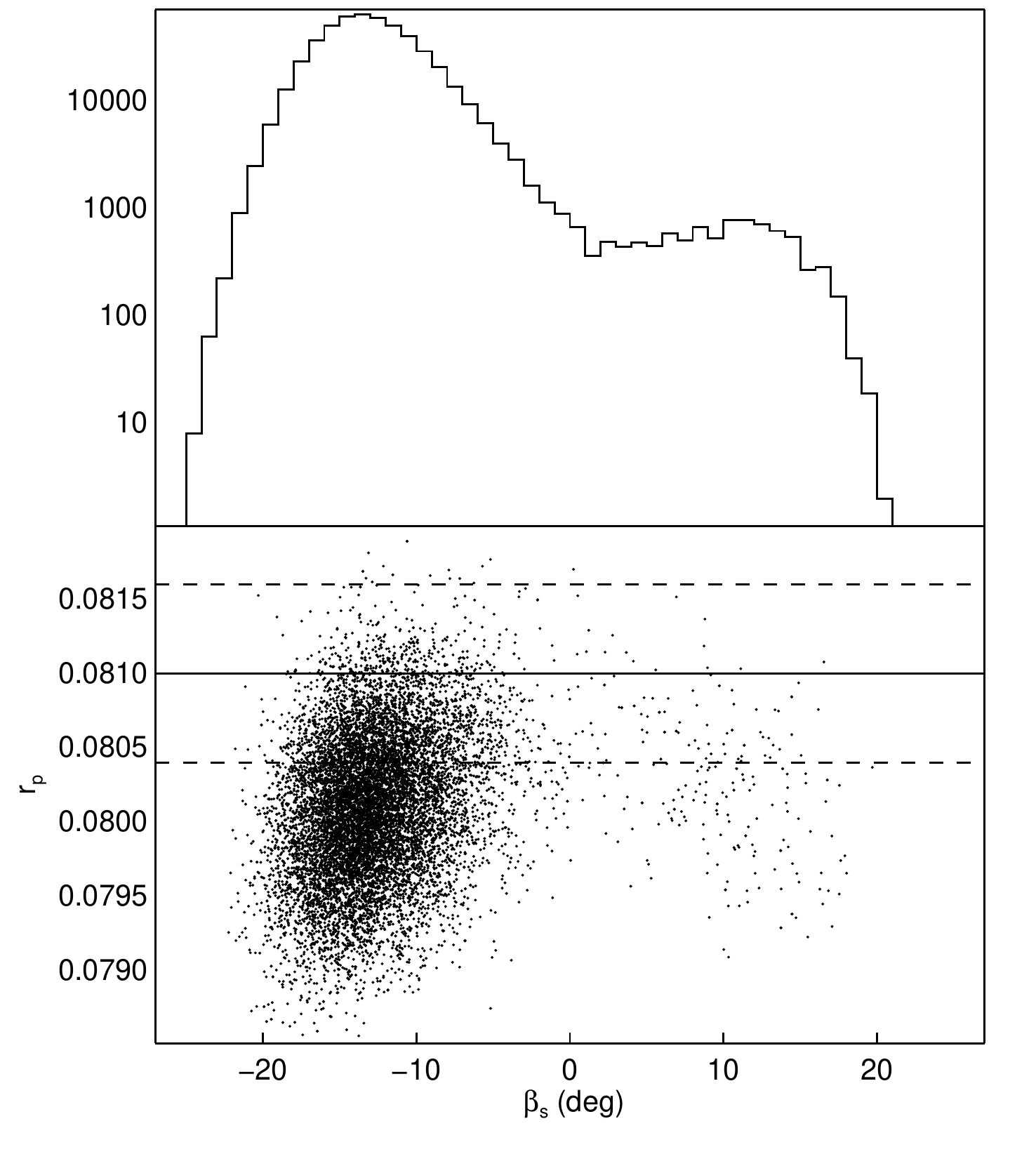}
\caption {\label{fig:beta_s} {\bf Projected obliquity of the secondary
    star.} {\it Upper panel:} a histogram of the posterior of
  $\beta_{\rm s}$ with logarithmic y-axis. In addition to the main peak
  around an angle of $-13^{\circ}$ there is a smaller peak at similar
  but positive values. {\it Lower panel:} a random sup-sample of the
  posterior in the $\beta_{\rm s}-r_{\rm p}$ plane. The  horizontal
  lines indicate the photometric prior and uncertainty interval for
  $r_{\rm p}$.  For larger values of $r_{\rm
    p}$ smaller absolute values of $\beta_{\rm s}$ are found.}
  \end{center}
\end{figure}

\subsection{Stellar Rotation and  Projected Obliquities}
\label{sec:rot}

The main result of our analysis is that the sky projections of the two
stellar rotation axes $\beta_{\rm p}=-1.8\pm1.6^{\circ}$ and
$\beta_{\rm s}=-13\pm4^{\circ}$ indicate close alignment between the
stellar rotation axes and the orbital angular momentum. However while
the value for $\beta_{\rm p}$ is consistent with prefect alignment $\beta_{\rm
  s}$ seems to indicate a small but significant misalignment. How
robust is this finding of a small misalignment? We note that we have
somewhat lower S/N observations during the secondary eclipse than
during the primary eclipse and fewer observations directly before,
during and after the eclipse (See Table~\ref{tab:log_nsv5783}, and
Figures~\ref{fig:nsv5783_primary}, \ref{fig:nsv5783_secondary}, and
\ref{fig:nsv5783_rvs}). To test the robustness of the result we reran
the MCMC chain with different model assumptions. For example we
constrained the model more, by leaving $\gamma_{\rm i}$ and the line
depths of both stars tied to each other, or we left $\zeta_{\rm i}$
and the limb darkening parameters completely free. We also excluded
some observations to test if a small number of observations are having
a disproportionate influence on the result. For all these runs we
found a negative $\beta_{\rm s}$. The result closest to alignment was
$\beta_{\rm s}=-9\pm5^{\circ}$. At the same time the result for
$\beta_{\rm p}$ did not change by more than $0.4^{\circ}$ during these
tests. There are, however, two peculiarities about our result for
$\beta_{\rm s}$. The posteriors for all the other parameters have only
a single peak, while the posterior of $\beta_{\rm s}$ has a small (two
orders of magnitude lower) secondary peak at positive angles around
$\beta_{\rm s}=13^{\circ}$ (Figure~\ref{fig:beta_s}, upper panel). In
addition there is a correlation between $\beta_{\rm s}$ and $r_{\rm
  p}$ (Figure~\ref{fig:beta_s}, lower panel). $r_{\rm p}$ is the only
parameter for which we do find a more than $1$--$\sigma$ displacement
between the prior constraints and results (Table~\ref{tab:results}).
Taken the above mentioned points into consideration we are confident
that $|\beta_{\rm s}|<17^{\circ}$ but we cannot exclude a small
projected obliquity for the secondary star, with the data at hand.

For the projected rotation speeds we find $v \sin i_{\rm
  p}=141.4\pm1.2$~km\,s$^{-1}$ and $v \sin i_{\rm
  s}=137.8\pm1.1$~km\,s$^{-1}$. We consider the formal uncertainties
in the $v \sin i_{\rm i}$ to be too low for the following reasons: We
tested different limb darkening laws and found for example that for a
linear limb darkening law the best fitting $v \sin i$ values are lower
by about $4$~km\,s$^{-1}$. Also we suspect that our particular choice
of parameterization of the stellar surface velocity fields will
influence the values we find for the projected rotation. We therefore
estimate that a uncertainty of $5$~km\,s$^{-1}$ is more realistic and
also indicated that uncertainty in Table~\ref{tab:results}. As
mentioned above, normalization for each observed spectrum is included
in our routine, hence any uncertainty in normalization is already
incorporated in the formal uncertainty.

\begin{figure}
  \begin{center} \includegraphics[width=8cm]{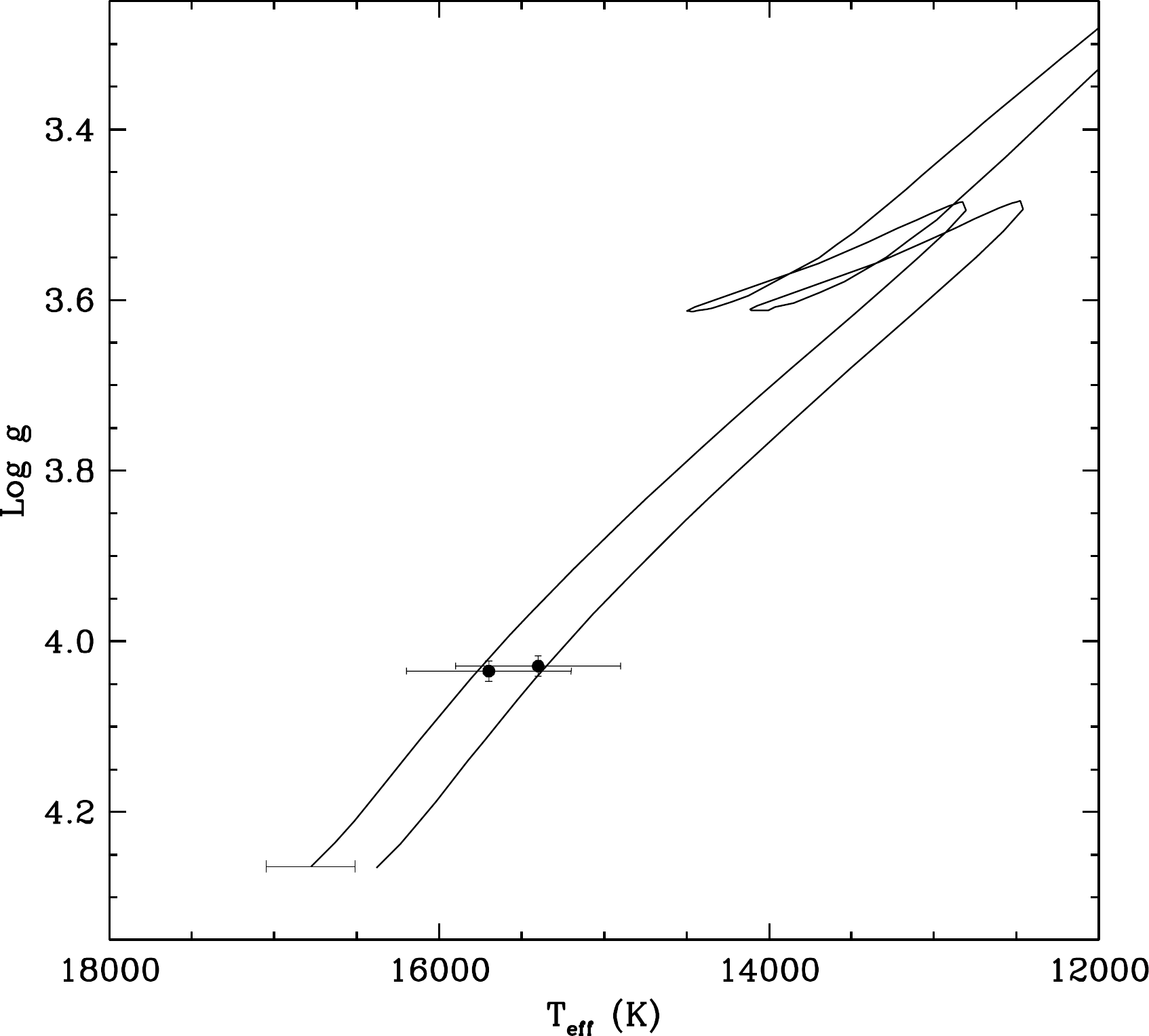} \caption
    {\label{fig:age} {\bf Stellar evolution tracks for EP\,Cru} from
      the Yonsei-Yale series \citep{yi2001,demarque2004} compared with
      the measurements. The tracks are interpolated to the measured
      masses and a metallicity of [Fe/H]~$ = +0.03$ that best fits the
      temperatures. The uncertainty in the location of the tracks that
      comes from the mass error is indicated with an error bar for the
      primary, and is similar for the secondary. The age according to
      these models is $57\pm5$~Myr.}
  \end{center}
\end{figure}

We note that the projected rotation speeds are similar to the average
rotation speed for B stars ($v \sin i= 130$~km\,s$^{-1}$), as analyzed
by \cite{abt2002}. However the stars might have undergone a change of
$v$ due to tidal interactions (Section~\ref{sec:tides}). Therefore we
can not conclude from the similarity of the measured $v \sin i$ to the
expected $v$ that $\sin i$ is close to unity. Nevertheless it seems
unlikely that the stars have large inclinations relative to the line of
sight and at the same time their projected axes on the plane of the
sky are both aligned. In what follows we assume that not only the
projections of the rotation axes are small, but that the axes
themselves are aligned too ($\sin i \approx 1$).

Concerning the values for the macro-turbulence, we expect that the
value we find does not have a simple physical interpretation. This is
because we assume equal brightness of raising and falling material as
well as equal surface coverage of movement tangential and radial to
the stellar surface, both assumptions do not need to be fulfilled in
reality. We did test if there is a strong dependence of the measured
values for projected obliquities on our adopted model for
macro-turbulence, and found none.

\begin{table}
 \caption{Expected Apsidal Motion in EP\,Cru }
 \label{tab:apsidal}
 \begin{center}
 \smallskip
     \begin{tabular}{l r@{$\pm$}l }
	\hline
	\hline
	\noalign{\smallskip}
        Expected apsidal motion & \multicolumn{2}{l}{(arcsec cycle$^-1$)} \\
	\noalign{\smallskip}
         \hline
       	\noalign{\smallskip}
        $\dot{\omega}_{\rm GR}$ & 1.87&0.023\\
         $\dot{\omega}_{\rm Tides}$ & 0.76&0.11 \\
         $\dot{\omega}_{\rm Rot}$  & 6.17&1.00 \\
         $\dot{\omega}_{\rm Total}$  & 8.8&1.1 \\
         \noalign{\smallskip}
         \hline
     \end{tabular}
     \tablecomments{ 
       $\dot{\omega}_{\rm GR}$ denotes apsidal motion due to General
       Relativity, $\dot{\omega}_{\rm Tides}$ due to tidal
       distortions, and $\dot{\omega}_{\rm Rot}$ due to rotational
       distortion.}
   \end{center} 
\end{table}

\subsection{Absolute Dimensions and Age}
\label{sec:dim}

From the posterior of our MCMC chain we find $K_{\rm
  p}=102.2\pm1.5$~km\,s$^{-1}$ and $K_{\rm
  s}=106.2\pm1.4$~km\,s$^{-1}$ in agreement with values from the
literature (Table~\ref{tab:results}). We also calculated the $K_{\rm
  i}$ values only using out of eclipse data, making them less
dependent on any assumption included in our eclipse model. With
$K_{\rm p}=101.9\pm1.5$~km\,s$^{-1}$ and $K_{\rm
  s}=106.2\pm1.6$~km\,s$^{-1}$ we obtain consistent
results.\footnote{John Southworth provided us with the 5 out of
  eclipse spectra used
  in the \cite{clausen2007} study and we found that these are
  consistent with our data set. Because of the potential small change
  in the argument of periastron over the last 20~years they have not been
  included in this study.}

With the new spectroscopic data we not only obtain precise mass
estimates for both stars but also improve on the absolute radii. This
is because the accurate scaled radii $r_{i}$ obtained via photometry
need to be multiplied by the absolute scale of the system, the
semi-major axis ($a$).

With the new values for the stellar masses, surface gravities ($\log
g_{i}$), and the effective temperatures $T_{\rm eff\,i}$ measured by
\citep[][see also Table~\ref{tab:nsv5783_overview}]{clausen2007} we
can estimate the stellar ages using stellar evolution models. Here we
employ the Yonsei-Yale evolutionary tracks
\citep{yi2001,demarque2004}. We find a good fit for solar metallicity
and an age of $57\pm5$~Myr (Figure~\ref{fig:age}). Another good check
is the temperature difference between the stars, since the difference
is probably better determined than the absolute temperatures. Indeed
the temperature difference predicted by the models (i.e., the
separation between the evolutionary tracks) is in good agreement with
the temperature difference measured by \cite{clausen2007}.

\subsection{Apsidal Motion}
\label{sec:apsidal}

Now that the stellar rotation is known we can calculate the expected
apsidal motion in the EP\,Cru system. We use the apsidal motion
constant $\log(k_{2})=-2.3$ from \cite{claret2004b} for both stars in
the system. We assign an uncertainty of $0.1$ in $\log$ space to this
constant. From the results in Table~\ref{tab:apsidal} we can see that
we expect a shift of $\approx 1.6\pm0.2^{\circ}$ over the last $20$
years (which approximately have elapsed since the photometric
measurements). Most of this shift is expected because of deformation
of the stars by their rotation. That we find a small increase in the
argument of the periastron and a ($\approx22$\,minutes) earlier primary
eclipse than expected from linear ephemeris seems to indicate apsidal
motion of the order of magnitude as expected. However spectroscopic
data is not very good at determining $\omega$ and we find it difficult
to estimate the significance of the measured value for $\omega$.
Therefore to make a meaningful comparison between the measured and
expected apsidal motion, new photometric eclipse timings should be
undertaken.

\begin{table}
 \caption{EP\,Cru and DI\,Her}
 \label{tab:compare}
 \begin{center}
 \smallskip
     \begin{tabular}{l r@{$\pm$}l r@{$\pm$}l }
	\hline
	\hline
	\noalign{\smallskip}
         Parameter &\multicolumn{2}{c}{ EP\,Cru}  &\multicolumn{2}{c}{DI\,Her} \\
        \noalign{\smallskip}
          \hline
          \noalign{\smallskip}
	Sp.\ Type & \multicolumn{2}{c}{B5V+ B5V} &  \multicolumn{2}{c}{B5V+ B5V\tablenotemark{$\star$}} \\
	$P$   (days) & \multicolumn{2}{c}{$11.08$} & \multicolumn{2}{c}{$10.55$\tablenotemark{$\star$}}  \\
        $e$ & $0.1874$&0.0005 & $0.489$&$0.003$\tablenotemark{$\star$} \\
        $M_{\rm p}$ ($M_{\odot}$) &  5.02&0.13  & 5.17&0.11\tablenotemark{$\star$}  \\
        $M_{\rm s}$ ($M_{\odot}$) &  4.83&0.13 &  4.52&0.07\tablenotemark{$\star$} \\
        $R_{\rm p} $ ($R_{\odot}$)  &  3.590&0.035 & 2.681&0.046\tablenotemark{$\star$} \\
        $R_{\rm s}$ ($R_{\odot}$)  &  3.495&0.034 & 2.478&0.046\tablenotemark{$\star$} \\     
        $\beta_{\rm p}$ ($^{\circ}$)  & $-1.8$&1.5 & 72&4\tablenotemark{$\dagger$} \\ 
        $\beta_{\rm s}$ ($^{\circ}$)  &  \multicolumn{2}{c}{$<17$} &  $-84$&8\tablenotemark{$\dagger$} \\ 
        $v \sin i_{\rm p}$ (km\,s$^{-1}$) & 141.4&5 & 108&4\tablenotemark{$\dagger$} \\
        $v \sin i_{\rm s}$ (km\,s$^{-1}$)  & 137.8&5 & 116&4\tablenotemark{$\dagger$} \\
        $v\,{\rm syn}_{\rm p}$ (km\,s$^{-1}$)  & 16.40&0.16 & 12.85&0.24 \\
        $v\,{\rm syn}_{\rm_s}$ (km\,s$^{-1}$)  & 16.04&0.15 & 11.89&0.24     \\
        $v\,{\rm ps}_{\rm p}$ (km\,s$^{-1}$) & 19.51&0.19 & 34.3&0.7 \\
        $v\,{\rm ps}_{\rm s}$ (km\,s$^{-1}$) & 19.00&0.18 & 31.8&0.7 \\
         Age (Myr)  &  57&5  & 4.5&2.5\tablenotemark{$\ddagger$} \\
	\noalign{\smallskip}
        \noalign{\smallskip}
	\hline
	\noalign{\smallskip}
        \noalign{\smallskip}
        $\star$Data from \cite{torres2010}\\
        $\dagger$Data from \cite{albrecht2009}\\
        $\ddagger$Data from \cite{claret2010}
     \end{tabular}
     \tablecomments{ 
       $v\,{\rm syn}_{\rm i}$ denotes the stellar rotation speed for an
       aligned star which rotation is synchronized with the orbital period.
       $v\,{\rm ps}_{\rm i}$ denotes the pseudo synchronized value as
       defined by \cite{hut1981}. To calculate $v\,{\rm ps}_{\rm i}$
       we used a $\Omega_{\rm ps}$ of $0.8$ \citep[see][Figure~3]{hut1981}.} 
   \end{center} 
\end{table}

\section{The Alignment in Context}
\label{sec:tides}

Having established the absolute dimensions, age, and state of rotation
in the EP\,Cru system we can now compare EP\,Cru to its apparently
younger sibling DI\,Her. In Table~\ref{tab:compare} we reprint some of
the values from EP\,Cru. According to these values the two systems are
similar, apart from two characteristics: 1) their ages, EP\,Cru is
about an order of magnitude older, 2) EP\,Cru appears to have aligned
axes which is definitely not the case for DI\,Her. We would like to
find a picture in which the misalignment in the young DI\,Her system
can be explained as well as the alignment in the older EP\,Cru system.

Knowing that the scaled radii are large enough in these systems to
allow for substantial tides, we might suspect that the difference in
spin-orbit alignment is a result of observing these systems at
different stages in their evolution rather than them having two
different formation and evolution paths. The hypothesis would be that
both stars had misaligned axes, and we see EP\,Cru with aligned axes
only because it is older and tides have had enough time to align the
axes.

The large eccentricities seen in both DI\,Her and EP\,Cru is
consistent with this hypothesis, because tides first align and
synchronize rotation and only on a longer timescale do they
circularize the orbit. This is mainly due to the higher amount of
angular momentum stored in the orbital motion compared to the stellar
rotation, and for systems with a low-mass secondary this is not
necessary the case. However another finding makes the hypothesis
difficult to reconcile with current tidal theories. The stars rotate
at $\sim9$ times the speeds expected for synchronized or
pseudosynchronized states (Table~\ref{tab:compare}). Thus tides have
not yet synchronized the stellar rotation speeds in the EP\,Cru
system. Formulations of tidal interactions predict that damping of any
significant spin-orbit misalignment should occur on the same time
scale as synchronization of the rotation
\citep{hut1981,eggleton2001}.\footnote{The timescales are not exactly
  equal, and which is faster depends on the ratio between the orbital
  and rotational angular momentum in the equilibrium states. For a
  system like EP\,Cru the timescale for pseudo synchronization is
  about twice the timescale for alignment \citep{hut1981}. } This is
because in these tidal models, a single coefficient describes the
coupling between tides and rotation. When stellar rotation is much
faster than the synchronized value rotation around any axis is damped
by about the same amount. Thus the angle between the overall angular
momentum and stellar spin does not change: only the rotation speed is
reduced. When the stellar rotation around a axis parallel to the
orbital angular momentum approaches the synchronized value than
rotation around this axis couples less to tides. Rotation around any
other axis is still damped by tides, which only ceases when the
rotation around these axes stops. The stellar rotation aligns to the
orbital axis.

\begin{figure}
  \begin{center}
    \includegraphics[width=8cm]{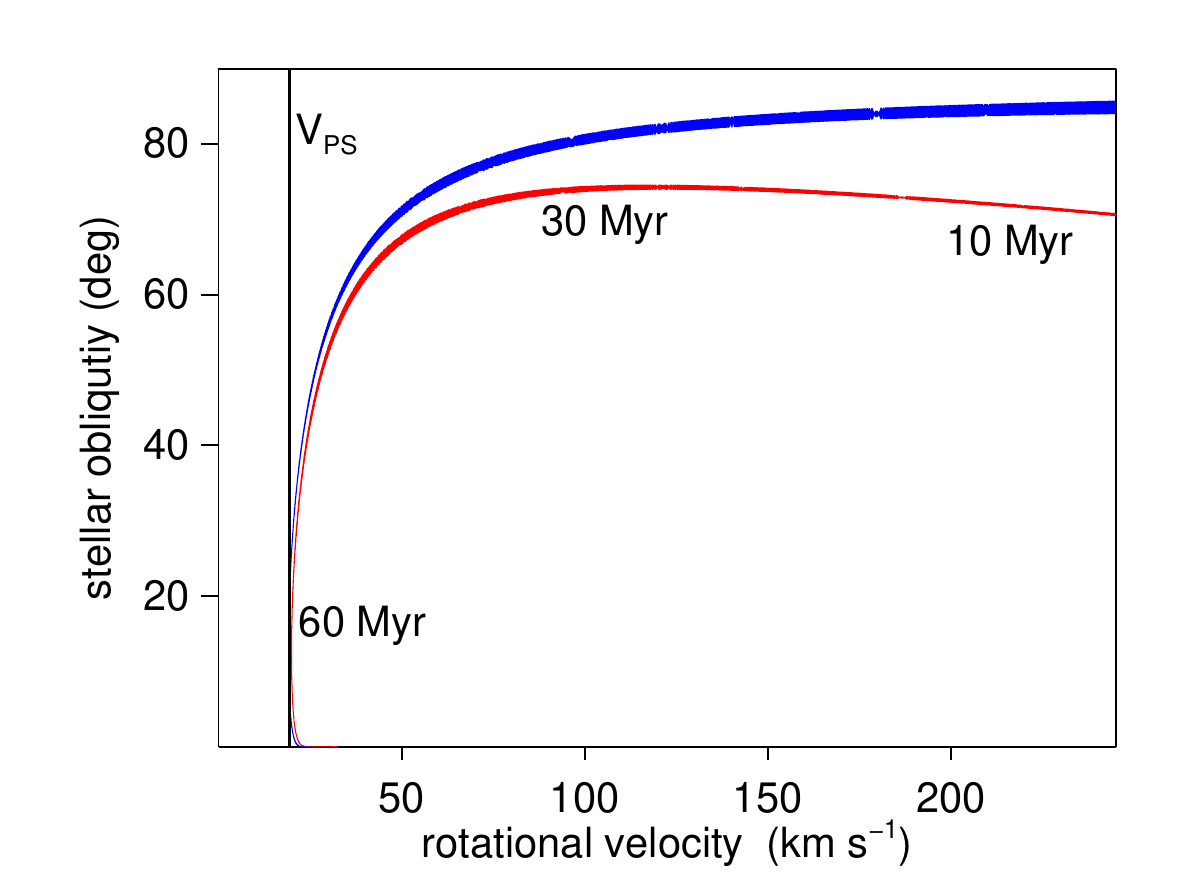}
    \caption {\label{fig:tides} {\bf Tidal evolution of a system
        similar to EP\,Cru, but with misaligned spin axes.} The blue
      and red lines show the evolutions of primary and secondary
      obliquities (angle between stellar spin and orbital plane) in a
      system with the parameters of EP\,Cru. However we started the
      run with obliquities which have been measured in the
      DI\,Her system and faster stellar rotation. We included the
      stellar evolution (in particular the change of the stellar radii
      with time) of the system as estimated with the Y$^{2}$-evolutionary
      tracks \citep{yi2001,demarque2004} and set the viscous time ($t_{\rm
      V}$) to 50\,000 years, about 1\,000 times larger than what is normally assumed for
    late type stars. (A lower value of $t_{\rm V}$ would lead to an
    overall faster tidal evolution but will leave the ratio of
    the  alignment and synchronization timescales unchanged.)  
    There is little evolution in the stellar obliquities until the
    rotation speeds approach the pseudo-synchronized value for
    rotation ($V_{\rm PS}$), which is indicated by the vertical line
    and is currently similar for the two stars (Table~\ref{tab:compare}). }
  \end{center}
\end{figure}

To illustrate this point we used the TOPPLE tidal-evolution code
developed by \cite{eggleton2001}. For this simulation we used the
EP\,Cru parameters from Table~\ref{tab:compare} but with initial
obliquities taken from DI\,Her, and an initial faster stellar rotation
speed at zero-age main sequence. The results are shown in
Figure~\ref{fig:tides}. The stellar obliquities remain large until the
stellar rotation speeds approach synchronization, at which point
obliquities are damped. This suggests that EP\,Cru had aligned axes
when it was as young as DI\,Her, implying in turn that DI\,Her and
EP\,Cru do not represent different stages of one evolution, but rather
two different evolution paths.

At the moment it is not possible to make more general statements as
only a few measurements of obliquities have been carried out in close
double-star systems. Furthermore most of these have been conducted in
Algol systems which have undergone mass transfer \citep[see Table~1
of][]{albrecht2011}. Obliquity observations should be carried out in a
variety of systems. Of particular interest would be young systems with
short orbital periods with and without a third star. The systems
should be young to minimize the influence of tides, they should have
orbital periods ranging from few days to few tens of days. Obliquity
measurements in these systems would be helpful in testing predictions
of KCTF and thereby of close binary formation. Measurements of
obliquities in wider systems would probe the length scale over which
the primordial angular momentum was influential. Conducting such
measurements is the aim of the BANANA project.

\section{Summary}
\label{sec:summary}

We have analyzed high resolution spectra of the eclipsing close double
star system EP\,Cru. We obtained absolute dimensions and
showed that the rotation axes of both stars are aligned with each
other and the orbital rotation ($\beta_{\rm p}= -1.8\pm1.6^{\circ}$
and $|\beta_{\rm s}|<17^{\circ}$). 

EP\,Cru is similar in its orbital and stellar characteristics to
DI\,Her. The two exceptions are that DI\,Her is younger and has two
strongly misaligned stellar rotation axes. We have been unable to show
that both systems represent different stages of one evolution
path. This is because the stars in EP\,Cru rotate at a few times their
synchronized value and tidal theory predicts that synchronization
occurs around the same time as alignment. Therefore the two systems
likely represent two different formation scenarios rather then two
different evolutionary stages. The sample of close double star systems
for which the obliquities are measured remains small. We plan to
ratify this situation by measuring obliquities in more close double
star systems in the framework of the BANANA project.

\acknowledgments 

The authors are grateful to Peter Eggleton for insightful discussions
on binary evolution and for making his TOPPLE code available to us. We
thank John Southworth for providing us with the spectra used in the
study by \cite{clausen2007}. S.A. acknowledges support during part of
this project by a Rubicon fellowship from the Netherlands Organisation
for Scientific Research (NWO). Work by S.A.\ and J.N.W.\ was supported
by NASA Origins award NNX09AB33G and NSF grant No.\ 1108595. D.C.F.
acknowledges NASA support through Hubble Fellowship grant
HF-51272.01-A, awarded by STScI, operated by AURA under contract NAS
5-26555. This research has made use of the following web resources:
{\tt simbad.u-strasbg.fr,
  adswww.harvard.edu,arxiv.org, http://arxiv.org}

\end{document}